\documentclass{article}

\usepackage{microtype}
\usepackage{graphicx}
\usepackage{subfigure}
\usepackage{booktabs} 
\usepackage{commath}
\usepackage{bbm}

\usepackage [english]{babel}
\usepackage [autostyle, english = american]{csquotes}
\MakeOuterQuote{"}

\makeatletter
\newcommand*{\indep}{%
  \mathbin{%
    \mathpalette{\@indep}{}%
  }%
}
\newcommand*{\nindep}{%
  \mathbin{
    \mathpalette{\@indep}{\not}
  }%
}
\newcommand*{\@indep}[2]{%
  \sbox0{$#1\perp\m@th$}
  \sbox2{$#1=$}
  \sbox4{$#1\vcenter{}$}
  \rlap{\copy0}
  \dimen@=\dimexpr\ht2-\ht4-.2pt\relax
  \kern\dimen@
  {#2}%
  \kern\dimen@
  \copy0 
} 

\newcommand{\mbp}{\mathbb{P}}
\newcommand{\mbe}{\mathbb{E}}

\usepackage{lipsum,graphicx,multicol}

\usepackage{mathtools}
\newcommand{\defeq}{\vcentcolon=}

\DeclarePairedDelimiter{\ceil}{\lceil}{\rceil}
\usepackage{amsmath,amsfonts,amssymb,amsthm,epsfig,epstopdf,titling,url,array}
\theoremstyle{plain}
\newtheorem{thm}{Theorem}[section]
\newtheorem{lem}[thm]{Lemma}
\newtheorem{prop}[thm]{Proposition}
\newtheorem{assump}[thm]{Assumptions}

\theoremstyle{definition}

\newtheorem{exmp}{Example}[section]

\theoremstyle{remark}

\usepackage{hyperref}

\usepackage[accepted]{icml2018}

\icmltitlerunning{Minimal I-MAP MCMC}

\begin{document}

\twocolumn[
\icmltitle{Minimal I-MAP MCMC \\for Scalable Structure Discovery in Causal DAG Models}



\icmlsetsymbol{equal}{*}

\begin{icmlauthorlist}
\icmlauthor{Raj Agrawal}{mit1,mit2,mit3}
\icmlauthor{Tamara Broderick}{mit1,mit2}
\icmlauthor{Caroline Uhler}{mit2,mit3}
\end{icmlauthorlist}

\icmlaffiliation{mit1}{Computer Science and Artificial Intelligence Laboratory}
\icmlaffiliation{mit2}{Institute for Data, Systems and Society}
\icmlaffiliation{mit3}{Laboratory for Information and Decision Systems, Massachusetts Institute of Technology}

\icmlcorrespondingauthor{Raj Agrawal}{r.agrawal@csail.mit.edu, r.agrawal.raj@gmail.com}

\icmlkeywords{Machine Learning, ICML}

\vskip 0.3in
]



\printAffiliationsAndNotice{}  

\begin{abstract}
Learning a Bayesian network (BN) from data can be useful for decision-making or discovering causal relationships. However, traditional methods often fail in modern applications, which exhibit a larger number of observed variables than data points. The resulting uncertainty about the underlying network as well as the desire to incorporate prior information recommend a Bayesian approach to learning the BN, but the highly combinatorial structure of BNs poses a striking challenge for inference. The current state-of-the-art methods such as order MCMC are faster than previous methods but prevent the use of many natural structural priors and still have running time exponential in the maximum indegree of the true directed acyclic graph (DAG) of the BN. We here propose an alternative posterior approximation based on the observation that, if we incorporate empirical conditional independence tests, we can focus on a high-probability DAG associated with each order of the vertices. We show that our method allows the desired flexibility in prior specification, removes timing dependence on the maximum indegree, and yields provably good posterior approximations; in addition, we show that it achieves superior accuracy, scalability, and sampler mixing on several datasets.
\end{abstract}

\section{Introduction}
\emph{Bayesian networks} (BNs)---or probabilistic graphical models based on directed acyclic graphs (DAGs)---form a powerful framework for representing complex dependencies among random variables. Learning BNs among observed variables from data has proven useful in decision tasks---such as credit assessment or automated medical diagnosis---as well as discovery tasks---such as learning gene regulatory networks \citep{causality_book, gene_expr_anal, pearl_causality, epidemiology_models,credit_risk}.
When the number of data points is much larger than the number of observed variables, a point estimate of the BN can be found using constraint-based or greedy search methods \citep{causality_book,chick2002,mmhc}.
However, in many applications, the number of observed variables is \emph{larger} than the number of data points. In this case, many BNs may agree with the observed data. 
A Bayesian approach offers a natural weighting scheme across BNs via the Bayesian posterior distribution. This weighting propagates coherently to point estimates and uncertainty quantification for structural features of the BN (such as the presence of certain directed edges). Moreover, a Bayesian approach allows the incorporation of prior information, which is common in applications of interest \cite{speed_priors}.

Unfortunately, due to the combinatorial explosion of the number of DAGs, exact posterior computation is intractable for graphs with more than thirty nodes \citep{exact_posterior, exact_struct}. This motivated \citet{struct_mcmc} to propose \emph{structure MCMC}, an approximate method using Markov chain Monte Carlo (MCMC). To overcome the slow mixing of structure MCMC and its variants \citep{edge_reversal}, 
\citet{Friedman2003} introduced \emph{order MCMC}. This algorithm achieves significantly faster mixing by running a Markov chain not over DAGs, but in the reduced space of permutations (i.e., orders) of the vertices of the DAG. 
However, order MCMC requires a particular (often undesirable) form for the prior on DAGs, and its iterations suffer from exponential time and memory complexity in the maximum indegree of the true DAG.
Heuristic fixes for scalability exist \citep{Friedman2003}, but their statistical costs are unclear.

In this paper, we propose a new method to leverage the improved mixing of MCMC moves in the permutation space; in addition, our approach comes with theoretical guarantees on approximation quality and allows more realistic DAG priors.
The key new ingredient is an observation by \citet{verma_pearl} that has been used for causal inference in the frequentist setting~\citep{gen_polyhedra, sp_algo}; namely, if we have access to conditional independence (CI) tests, we can associate each permutation with a unique DAG known as the \emph{minimal I-MAP} (independence map). This is the sparsest DAG that is consistent with a given permutation and Markov to a given set of CI relations. We prove that the vast majority of posterior mass is concentrated on the corresponding reduced space of DAGs, and we call our method \emph{minimal I-MAP MCMC}.

We start in Section~\ref{prelims} by reviewing BNs and Bayesian learning of BNs. We show how to reduce to the space of minimal I-MAPS in Section \ref{min_imap_space} and theoretically bound the posterior approximation error induced by this reduction. In Section~\ref{gen_model}, we show by an \emph{empirical Bayes} argument that sufficiently accurate CI tests allow using what amounts to our original prior and likelihood on DAGs but, crucially, restricted to the space of minimal I-MAPs. Thus, we demonstrate that our method allows arbitrary prior structural information. In Section~\ref{transposition_sec}, we present an MCMC approach for sampling according to this minimal I-MAP model and provide intuition for why it exhibits good mixing properties. Moreover, we prove that, for $p$ the number of observed variables and $k$ the maximum indegree of the true DAG, our method takes $\mathcal{O}(p^2)$ memory and $\mathcal{O}(p^4)$ time per MCMC iteration (vs.\ $\mathcal{O}(p^{k+1})$ time and memory for order MCMC). In Section \ref{sec:experiments} we empirically compare our model to order MCMC and \emph{partition MCMC}~\citep{partition_mcmc}, the state-of-the-art version of structure MCMC. In experiments we observe $\mathcal{O}(p^3)$ time scaling for our method, and we demonstrate better mixing and ROC performance for our method on several datasets.

\section{Preliminaries and Related Work} \label{prelims}
\subsection{Bayesian Networks}
Let $G = ([p], A)$ be a \emph{directed acyclic graph} (DAG) consisting of a collection of vertices $[p] := \{1,\ldots,p\}$ and a collection of arrows (i.e., directed edges) $A$, where $(i,j)\in A$ represents the arrow $i\to j$. A DAG induces a \emph{partial order} $\precsim$ on the vertices $[p]$ through $(i,j) \in A$ if and only if $i \precsim j$. Let $S_p$ be the symmetric group of order $p$. A \emph{topological order} of a DAG is a permutation $\pi\in S_p$ such that for every edge $(i,j) \in A$, $i\precsim j$ in $\pi$; thus it is a \emph{total order} that extends (i.e., is consistent with) the partial order of the DAG, also known as a \emph{linear extension} of the partial order. 

A \emph{Bayesian network} is specified by a DAG $G$ and a corresponding set of edge weights $\theta\in\mathbb{R}^{|A|}$. Each node in $G$ is associated with a random variable $X_i$. Under the \textit{Markov Assumption}, which we assume throughout, each variable $X_i$ is conditionally independent of its nondescendants given its parents, i.e., the joint distribution factors as
$\prod_{i=1}^p \mathbb{P}\big(X_i \ | \ \mathsf{Pa}_G(X_i)\big),$ where $\mathsf{Pa}_G(X_i)$ denotes the parents of node $X_i$ \citep[Chapter 4]{causality_book}. This factorization implies a set of \textit{conditional independence} (CI) relations that can be read off from the DAG $G$ by \emph{d-separation}. The \textit{faithfulness assumption} states that the only CI relations realized by $\mathbb{P}$ are those implied by d-separation in $G$ \citep[Chapter 4]{causality_book}. DAGs that share the same d-separation statements make up the \textit{Markov equivalence class} of a DAG \citep[Chapter 3]{lauritzen_book}. The Markov equivalence class of a DAG can be uniquely represented by its \textit{CP-DAG}, which places arrows on those edges consistent across the equivalence class \cite{cpdag}. The arrows of the CP-DAG are called \textit{compelled edges} and represent direct causal effects \cite{cpdag}.



\subsection{Bayesian Inference for DAG models}

In many applications, the goal is to recover a function $f(G)$ of the underlying causal DAG $G$ given $n$ i.i.d.~samples on the nodes, which we denote by $D = \{X_{mi}: m\in [n], i\in [p]\}$. 
For example, we might ask whether a directed edge $(i,j)$ is in $A$, 
or we might wish to discover which nodes are in the Markov blanket of a node $i$. 
In applications where $n$ is large relative to $p$, a point estimate of $G$---and thereby of $f(G)$---suffices from both a practical and theoretical perspective \cite{chick2002}. However, in many applications of modern interest, $n$ is small relative to $p$. 
In this case there may be many DAGs that agree with the observed data and it is then desirable to infer a distribution across DAGs instead of outputting just one DAG. 
Taking a Bayesian approach we can define a \emph{prior} $\mathbb{P}(G)$ on the space of DAGs, which can encode prior structural knowledge about the underlying DAG---as well as desirable properties such as sparsity.
The \emph{likelihood} $\mbp(D \mid G)$ is obtained by marginalizing out $\theta$:
\begin{align*}
	\mbp(D \mid G) &= \int_{\theta} \mbp( D, \theta \mid G) \; d\theta \\
    	&= \int_{\theta} \mbp(D \mid \theta, G) \mbp(\theta \mid G) \; d\theta
\end{align*}
and can be tractably computed for certain classes of distributions \cite{bge_orig,bge_score}. Applying Bayes theorem yields the \emph{posterior distribution} $\mbp(G \mid D) \propto \mbp(D \mid G) \mbp(G)$, which describes the state of knowledge about $G$ after observing the data $D$. From the posterior one can then compute 
$\mbe_{\mbp(G \mid D)} f(G)$, the posterior mean of the function of interest. Note that in the common setting where $f$ takes the form of an indicator function, this quantity is simply a posterior probability.  

Unfortunately, computing the normalizing constant of the posterior distribution is intractable already for moderately sized graphs, since it requires evaluating a sum over the space of all DAGs on $p$ vertices 
\cite{exact_posterior}. To sample from the posterior without computing the normalizing constant, \citet{struct_mcmc} proposed \emph{structure MCMC}, which constructs a Markov chain on the space of DAGs with stationary distribution equal to the exact posterior. $T$ samples $\{G_t\}$ from such a Markov chain can then be used to approximate the posterior mean of the function of interest, namely $\mbe_{\mbp(G \mid D)} f(G) \approx T^{-1} \sum_{t=1}^{T} f(G_t)$.

Problematically, the posterior over DAGs is known to exhibit many local maxima, so structure MCMC exhibits poor mixing even on moderately sized problems \cite{Friedman2003,ellis_wong}. To overcome this limitation, \citet{Friedman2003} proposed \emph{order MCMC}, which constructs a Markov chain on the space of permutations, where the moves are transpositions.  The posterior over orders is smoother than the posterior over DAGs, since the likelihood corresponding to each order is a sum over many DAGs, and increased smoothness usually leads to better mixing behavior. However, strong modularity assumptions are needed to make computing the likelihood tractable. Even under these assumptions, there remains a considerable computational cost: namely, let $k$ be the maximum indegree of the underlying DAG, then the likelihood can be computed in $\mathcal{O}(p^{k+1})$ time and memory \cite{Friedman2003}. Hence, in practice $k$ can be at most 3 or 4 for this method to scale to large networks. The Monte Carlo estimate $\frac{1}{T}\sum_{i=1}^T f(G_{\pi_t})$, where $G_{\pi_t}$ is drawn from $\mathbb{P}(G \mid \pi_t, D)$ and $\pi_t$ is sampled from the posterior over permutations $\mathbb{P}(\pi \mid D)$, is then used to approximate the posterior mean of the function of interest. \citet{Friedman2003} obtain a practical MCMC sampler when the prior over permutations is uniform, but such a model introduces significant bias toward DAGs that are consistent with more permutations \citep{ellis_wong}. Correcting for this bias by re-weighting each sampled DAG by the inverse number of its linear extensions can be done, but it is $\#P$ in general \cite{ellis_wong}.

A recent extension of order MCMC is \textit{partial order MCMC} \cite{partial_order}. This method works on the reduced space of partial orders, thereby leading to improved mixing as compared to order MCMC, but with a similar runtime. \citet{partition_mcmc} further introduced a related method known as \emph{partition MCMC}, which avoids the bias of order MCMC by working on the larger space of node partitions consisting of permutations and corresponding \emph{partition} elements. Although partition MCMC generally mixes more slowly than order MCMC, it was empirically found to mix more quickly than structure MCMC \cite{partition_mcmc}. 

\section{Reduction to the Space of Minimal I-MAPs} \label{min_imap_space}
To overcome the computational bottleneck of order MCMC and at the same time avoid the slow mixing of structure MCMC, we propose to restrict our focus to a carefully chosen, reduced space of DAGs that is in near one-to-one correspondence with the space of permutations. We construct this subspace of DAGs from the CI relations that hold in the data $D$. In Appendix \ref{A:ci_tests} we review a CI testing framework based on partial correlations for the Gaussian setting. 

Given a CI test, let $\hat{\mathcal{O}}^{(n)}_{i,j|S}(D,\alpha)$ be 1 if the corresponding CI test at level $\alpha$ based on the $n$ data points in $D$ was rejected---i.e., $X_i{\not\!\perp\!\!\!\!\perp} X_j\mid X_S$---and 0 otherwise. Let $\hat{\mathcal{O}}_n(D,\alpha)$ denote the collection of CI test outcomes across all triples $(i, j, S)$. Given $\hat{\mathcal{O}}_n(D,\alpha)$ we associate to each permutation $\pi\in S_p$ its \emph{minimal I-MAP} $\hat{G}_{\pi}$: a DAG with vertices $[p]$ and arrows $\pi(i) \rightarrow \pi(j)$ if and only if $\hat{\mathcal{O}}^{(n)}_{i,j|S}(D,\alpha) = 1$ with $i<j$ and $S = \{\pi(1) \cdots \pi(j-1) \} \setminus \{ \pi(i) \}$. In light of Occam's razor it is natural to consider this mapping, since removing any edge in $\hat{G}_{\pi}$ induces a CI relation that is not in $\hat{\mathcal{O}}_n(D,\alpha)$~\cite{causality_book,sp_algo}. 

Let $\hat{\mathcal{G}}:= \{\hat{G}_{\pi}\mid \pi\in S_p\}$. Then any posterior $\mathbb{P}(\pi\mid D)$ defined by a likelihood and prior on $S_p$ induces a distribution on the space of all DAGs, denoted by $\mathcal{G}$, namely
\begin{equation} \label{eq:imap_post_approx}
\hat{\mbp}( G \mid D) \defeq \sum_{\pi \in S_{p}} \mathbbm{1}\{ G = \hat{G}_{\pi}\} \mbp(\pi \mid D).
\end{equation}
This distribution places mass only on $\hat{\mathcal{G}}$ and weights each minimal I-MAP according to the posterior probability of sampling a permutation that is consistent with it. 

In the remainder of this section we introduce our main result showing that the posterior mean of a function based on the posterior $\mathbb{P}(G\mid D)$ defined by a likelihood and prior on $\mathcal{G}$ is well approximated by the posterior mean of the function based on the distribution $\hat{\mathbb{P}}(G\mid D)$ that has support only on $\hat{\mathcal{G}}$. Before stating our main result, we introduce the assumptions required for this result.

\begin{assump}\label{assumptions}
Let $(G^*,\theta^*)$ define the true but unknown Bayesian network. Let $\mathcal{O}^{*}$ be the equivalent of $\hat{\mathcal{O}}_n(D,\alpha)$ based on the true but unknown joint distribution on $G^*$. For each $\pi\in S_p$ let $G^*_\pi$ denote the minimal I-MAP with respect to $\mathcal{O}^{*}$. We make the following assumptions:
\vspace{-0.2cm}
\begin{enumerate}
\item[(a)] $X_{i} \mid (G^*,\theta^*)$ is multivariate Gaussian. \label{assum:a}
\vspace{-0.1cm}
\item[(b)] Let $\rho^*_{i, j \mid S}$ be the partial correlation derived from the Bayesian network $(G^*, \theta^*)$ for the triple $(i,j,S)$ and let $Q^{*} \defeq \sup_{(i,j,S)} \{|\rho^*_{i, j \mid S}|\}$. Then there exists $q^{*} < 1$ such that $\mbp(Q^{*} <  q^{*}) = 1$.
\vspace{-0.1cm}
\item[(c)] Let $R^{*} \defeq \inf_{(i,j,S)} \{|\rho^*_{i, j \mid S}| : \rho^*_{i, j \mid S}\neq 0\}$. Then there exists $r^{*} > 0$ such that $\mbp(R^{*} >  r^{*}) = 1.$
\vspace{-0.1cm}
\item[(d)] $\hat{G}_{\pi}$ is a sufficient statistic for $\mbp(G \mid \pi, D)$, i.e., $\mbp(G \mid \pi, D) = \mbp(G \mid \hat{G}_{\pi})$.
\vspace{-0.1cm}
\item[(e)] Let $A_\pi$ denote the event that $\{\hat{G}_{\pi} = G^*_{\pi}\}$. Then $\mbp(A_\pi \mid \hat{G}_{\pi}) = \mbp(A_\pi)$. 
\item[(f)] There exists some $M < \infty$ such that $\max_{G} |f(G)| \leq M$.
\end{enumerate}
\end{assump}
\vspace{-0.1cm}

An in-depth discussion of these assumptions is provided in Appendix \ref{dis_of_assump}. Assumption \ref{assum:a}(c) can be regarded as the Bayesian analogue of the \textit{strong-faithfulness assumption}, which is known to be restrictive~\cite{geometry_strong_faith} but is a standard assumption in causal inference for obtaining theoretical guarantees \cite{high_dim_PC, unif_consis_strong_faith}. Practitioners often choose $f$ to be an indicator function (e.g.~the presence of a directed edge), so Assumption \ref{assumptions}(f) is typically satisfied in practice.   

We now state our main result that motivates constructing a Markov chain on the reduced DAG space $\hat{\mathcal{G}}$ instead of $\mathcal{G}$.

\begin{thm} \label{thm:good_approx_feats}
Under Assumptions~\ref{assumptions}(a)-(f) it holds that
$$\left|\mbe_{\mbp(G \mid D)} f(G) - \mbe_{\hat{\mbp}( G \mid D)} f(G)\right| \leq 2f(n, p),$$
where $f(n, p) = C_1Mp^2(n - p)\exp \{{-C_2(r^{*})^2(n - p)} \}$.
\end{thm}
Theorem \ref{thm:good_approx_feats} is proven in Appendix \ref{pf:good_approx_feats}. The main ingredient of the proof is the following lemma that bounds the probability of the events $A_\pi^C$ for all $\pi$. 

\begin{lem} \label{thm:exp_concentrationv2}
Under Assumptions~\ref{assumptions} (a)-(c) there exist constants $C_1, C_2$ that depend only on $q^{*}$ such that
$$\mathbb{P}(G_{\pi}^{*} \neq \hat{G}_{\pi}) \leq  f(n,p), $$
for all $\pi\in S_p$, where $f(n,p)$ is as in Theorem~\ref{thm:good_approx_feats} and $\hat{G}_{\pi}$ is constructed using Fisher's z-transform to do CI testing at level $\alpha = 2\big(1 - \Phi\big(\frac{\sqrt{n}r^{*}}{2}\big)\big)$.
\end{lem}

From Theorem~\ref{thm:good_approx_feats} and Equation (\ref{eq:imap_post_approx}), it follows that
\begin{align}
	\nonumber
	\mbe_{\mbp(G \mid D)} [f(G)] &\quad\approx\quad \mbe_{\hat{\mbp}(G \mid D)} [f(G)] \\
    	\label{eq:min_imap_redux}
        &\quad=\quad \sum_{\pi \in S_{p}} f(\hat{G}_{\pi}) \mbp(\pi \mid D).
\end{align}
Hence, using the near one-to-one mapping between $S_p$ and $\hat{\mathcal{G}}$ to associate to each permutation a particular DAG, we can show that the posterior mean $\mbe_{\mbp(G \mid D)} f(G)$ can be well approximated by sampling  from a posterior over permutations. This is of particular interest given the observation by \citet{Friedman2003} that a posterior over permutations is generally smoother than a posterior over DAGs and hence more conducive to fast mixing in MCMC methods.  

\section{Bayesian Inference on Minimal I-MAPs} \label{gen_model}

Our original Bayesian generative model consisted of a prior $\mbp(G)$ and a likelihood $\mbp(D | G)$. In some sense, $\pi$ may be thought of an auxiliary random variable that aids our reduction to the minimal I-MAP space. But inventing a prior and likelihood for $\pi$ in order to arrive at the posterior $\mbp(\pi \mid D)$ in Equation~(\ref{eq:min_imap_redux}) may be conceptually difficult. In particular, it is natural to imagine we might have prior and modeling information for $G$ rather than $\pi$ in applications. And $S_p$ does not induce a partition in $\mathcal{G}$ \cite{ellis_wong}; see also Appendix \ref{dis:priors_topo_diff}. In this section, we demonstrate that, when the available CI information is sufficiently reliable, a good approximation to $\mbe_{\mbp(G \mid D)}[f(G)]$ can be obtained as follows.
\begin{align}
	\label{eq:bayes_min_imap}
	\mbe_{\mbp(G \mid D)}[f(G)] &\approx \sum_{\hat{G} \in \hat{\mathcal{G}}} f(\hat{G}) \mbp(\hat{G} | D), \textrm{ \quad where} \\
    \nonumber
    \mbp(\hat{G} | D) &\propto \mbp(D | G = \hat{G}) \mbp(G = \hat{G})
\end{align}
and the final two terms are the original likelihood $\mbp(D \mid G)$ and prior $\mbp(G)$ restricted to the minimal I-MAP space. This formula is intuitively appealing; it effectively says that we can obtain a good approximation of the desired posterior expectation by  simply restricting our original model to the minimal I-MAP space.

To show this, we start from Equation~(\ref{eq:min_imap_redux}) and let $\hat{\mathcal{O}}_{n} := \hat{\mathcal{O}}_{n}(D,\alpha)$ for brevity.
Note that $\mbp( \pi | D ) = \mbp( \pi | D, \hat{\mathcal{O}}_{n})$ since $\hat{\mathcal{O}}_{n}$ is a function of $D$.
Then, by Bayes theorem, 
\begin{equation} \label{eq:empirical_bayes}
	\mbp( \pi \mid D ) \propto \mbp( D \mid \pi, \hat{\mathcal{O}}_{n}) \mbp( \pi \mid \hat{\mathcal{O}}_{n}).
\end{equation}
Conditioning on a statistic of the data, namely $\hat{\mathcal{O}}_{n}$ here, before applying Bayes theorem may be thought of as an \emph{empirical Bayes} procedure \cite{darnieder_dep_priors}.

We examine each of the two factors on the righthand side of Equation~(\ref{eq:empirical_bayes}) in turn. Recall that $A_{\pi} := \{\hat{G}_{\pi} = G^{*}_{\pi}\}$ is the event that we make no CI errors.
First, note that
\begin{align}
	\nonumber
	\mbp( D \mid \pi, \hat{\mathcal{O}}_{n} )
    &= \sum_{G \in \mathcal{G}} \mbp( D \mid \pi, \hat{\mathcal{O}}_{n}, G ) \mbp( G \mid \pi, \mathcal{O}_{n}) \\
    \label{eq:G_expand}
    &= \sum_{G \in \mathcal{G}} \mbp (D \mid G) \mbp( G \mid \hat{G}_{\pi}).
\end{align}
Moreover, note that 
\begin{align*}
	\lefteqn{ \mbp( G \mid \hat{G}_{\pi}) } \\
     &= \mbp( G \mid \hat{G}_{\pi}, A_{\pi}) \mbp(A_{\pi} | \hat{G}_{\pi}) + \mbp( G \mid \hat{G}_{\pi}, A^{C}_{\pi}) \mbp(A^{C}_{\pi} | \hat{G}_{\pi}) \\
     &= \mbp( G \mid \hat{G}_{\pi}, A_{\pi}) \mbp(A_{\pi}) + \mbp( G \mid \hat{G}_{\pi}, A^{C}_{\pi}) \mbp(A^{C}_{\pi})
\end{align*}
By Assumption \ref{assumptions}(e), $\mbp(A_{\pi} | \hat{G}_{\pi}) = \mbp(A_{\pi})$. By Lemma \ref{thm:exp_concentrationv2}, $\mbp(A_{\pi})$ approaches 1 exponentially fast in $n$, and so $ \mbp(A_{\pi}^{C})$ approaches zero exponentially fast in $n$. Observing that
$\mbp( G | \hat{G}_{\pi}, A_{\pi}) = \mathbbm{1}\{ G = \hat{G}_{\pi}\}$ and that $\mbp( G | \hat{G}_{\pi}, A^{C}_{\pi})$ is bounded by one, we find $\mbp( G | \hat{G}_{\pi}) \approx \mathbbm{1}\{ G = \hat{G}_{\pi}\}$ for a sufficiently accurate CI test. Therefore, substituting back into Equation~(\ref{eq:G_expand}), we find that
$$
	\mbp( D \mid \pi, \hat{\mathcal{O}}_{n}) \approx \mbp (D \mid G = \hat{G}_{\pi}),
$$
the likelihood restricted to the space of minimal I-MAPs.

A similar argument, detailed in Appendix \ref{dis:prior_red}, yields that the second term in Equation~(\ref{eq:empirical_bayes}) is approximately equal to the prior restricted to the space of minimal I-MAPs:
$$
\mbp( \pi \mid \hat{\mathcal{O}}_{n})
	\approx \mbp(G = \hat{G}_{\pi}).
$$

Finally, if we let $\mbp(\hat{G}_{\pi} | D)$ represent the distribution over $\hat{G}_{\pi}$ proportional to the likelihood $\mbp (D | G = \hat{G}_{\pi})$ times the prior $\mbp(G = \hat{G}_{\pi})$, we can replace Equation~(\ref{eq:min_imap_redux}) with
Equation~(\ref{eq:bayes_min_imap}) at the beginning of this section, as was our goal.
In the next section we develop a Markov Chain Monte Carlo sampler with the desired stationary distribution, $\mbp(\hat{G}_{\pi} | D)$.

\section{Minimal I-MAP MCMC}
\label{transposition_sec}

In this section we develop a Markov Chain Monte Carlo sampler, which we call \emph{minimal I-MAP MCMC}, to generate approximate samples from the target distribution, $\mbp(\hat{G}_{\pi} | D)$. We show that unlike structure MCMC our approach is amenable to fast mixing. Furthermore, we show that minimal I-MAP MCMC overcomes the computational limitations of order MCMC, since its complexity does not depend on the maximum indegree of the underlying DAG $G^*$. 

Our minimal I-MAP MCMC algorithm is detailed in Algorithm~\ref{algo} for the Gaussian setting. Algorithm~\ref{update_min_imap}, denoted as \emph{update minimal I-MAP (UMI)}, is used as a step in Algorithm~\ref{algo} and describes how to compute a minimal I-MAP $\hat{G}_{\tau}$ from a minimal I-MAP $\hat{G}_\pi$ when $\pi$ and $\tau$ differ by an adjacent transposition without recomputing all edges; see also \citet{greedy_sp}. We prove the following proposition about the correctness of our sampler in Appendix \ref{pf:stat_dist}.

\begin{prop} \label{prop:stationary_distr}
In the Gaussian setting, the transitions in Algorithm~\ref{algo} define an ergodic, aperiodic Markov chain on $\mathcal{\hat{G}}$ with stationary distribution $\mbp(\hat{G}_{\pi} | D)$.
\end{prop}

Note that minimal I-MAP MCMC can easily be extended to the non-Gaussian setting by replacing the CI tests based on partial correlations by CI tests based on mutual information. However, for non-Gaussian data our theoretical guarantees do not necessarily hold.
 
In Section \ref{sec:experiments} we show empirically that minimal I-MAP MCMC mixes faster than other MCMC samplers. The following example provides intuition for this behavior.

\begin{exmp}
Suppose the true DAG $G^*$ is the star graph with arrows $2\to 1, 3\to 1, \dots p\to 1$. For the sake of simplicity, suppose $\mathcal{O}_n(D,\alpha) =\mathcal{O}^{*}$. Then for the permutation $\tau = (1 3 \cdots p 2)$ the corresponding minimal I-MAP $\hat{G}_{\tau}$ equals the fully connected graph. However, a single transposition from $\tau$ yields the permutation $\pi = (2 3 \cdots p 1)$, which is consistent with the DAG $G^*$. Hence minimal I-MAP MCMC can move in a single step from the fully connected graph to the correct DAG, while structure MCMC, which updates one edge at a time, would require many steps and could get stuck along the way. 
\end{exmp}
While this example is clearly idealized, it captures the intuition that traversing the space of minimal I-MAPs via transpositions allows the sampler to make large jumps in DAG space, which allows it  to escape local maxima faster and hence mix faster than structure MCMC. In the following result we characterize the memory and time complexity of minimal I-MAP MCMC, showing that unlike order MCMC it does not depend on the maximum indegree of the true DAG $G^*$. The proof is given in Appendix \ref{pf:time_mem_scale}.





\begin{prop} \label{prop:time_memory_ergodic}
Let $\kappa$ be the thinning rate of the Markov chain and $T$ the number of iterations. Consider minimal I-MAP MCMC (Algorithm \ref{algo}) with a proposal  distribution that puts mass only on adjacent transpositions, i.e.
\[
  q(\pi_t \rightarrow \pi_{t+1}) =
  \begin{cases}
                                   s & \text{if $\pi_t = \pi_{t+1}$} \\
                                   \frac{1-s}{p} & \text{if $I(\pi_t, \pi_{t+1}) = 1$} \\
  0 & \text{otherwise,}
  \end{cases}
\]
where $0<s<1$ and $I(\cdot, \cdot)=1$ if the permutations differ by a single adjacent transposition. 
This algorithm takes $\mathcal{O}(\kappa T p^2)$ memory and has average time complexity of $\mathcal{O}(Tp^4 + p^5)$.
Note that a transposition between the first and last element of a permutation is still considered an adjacent transposition in our definition. 
\label{bsp_time_complexity}
\end{prop}

Using a proposal that considers only adjacent transpositions leads to a considerable speed up. In particular, if we consider any possible transition, updating  $\hat{G}_{\pi_{t}}$ to $\hat{G}_{\pi_{t+1}}$ requires $\mathcal{O}(p^2)$ CI tests in general. But the cost is reduced to $\Theta(p)$ CI tests for adjacent transpositions that do not swap the first and last elements. Since performing a CI test based on partial correlations takes $\mathcal{O}(p^3)$ time \cite{part_cor_form}, this yields a total speed up of a factor of $p$ at each step. We should note that Algorithm \ref{algo} can be sped up by considering only adjacent transpositions that are connected by an edge; i.e., in minimal I-MAP space $\hat{\mathcal{G}}$ these adjacent transpositions would correspond to considering only \emph{covered edge} flips \cite{causality_book, greedy_sp}. 

We now comment on why our method does not face the computational intractability of order MCMC. 
Working in the space of minimal I-MAPS parametrized by permutations is similar in spirit to order MCMC, but our approximation of the posterior (that is, the approximation we make even before applying MCMC) allows us to avoid the poor scaling of order MCMC. In particular, the intractability of order MCMC arises due to the focus on an exact likelihood; acquiring this likelihood requires summing over $\mathcal{O}(p^{k+1})$ parent sets in order to sum over the full space of DAGs. In our case, we instead exploit the fact that the likelihood concentrates around a single DAG $\hat{G}_{\pi}$ once we condition on $\hat{\mathcal{O}}_n(D,\alpha)$. 

\begin{algorithm}[tb] 
   \caption{Minimal I-MAP MCMC}
\begin{algorithmic}
   \STATE {\bfseries Input:} \ Data $D$, number of iterations $T$, significance level $\alpha$, initial permutation $\pi_0$, sparsity strength $\gamma$, thinning rate $\kappa$
   \STATE {\bfseries Output:} $\hat{G}_{\pi_1}, \cdots, \hat{G}_{\pi_{\ceil{\kappa T}}}$
   \STATE Construct $\hat{G}_{\pi_0}$ from $\hat{\mathcal{O}}_n(D,\alpha)$ via Fisher's z-transform
   \FOR{$i=1$ {\bfseries to} $T$}
   \STATE Sample $\pi_i \sim q(\pi_{i-1} \rightarrow \pi_{i})$
   \STATE $\hat{G}_{\pi_i} = \textrm{UMI}(\pi_i, \pi_{i-1}, \hat{G}_{\pi_{i-1}}, \alpha, D)$ (Algorithm \ref{update_min_imap}, Appendix \ref{A:update_algo})
   \STATE $p_{i-1} = \log \mathbb{P}(D \mid \hat{G}_{\pi_{i-1}})\mbp(\hat{G}_{\pi_{i-1}})$
   \STATE $p_{i} = \log \mathbb{P}(D \mid \hat{G}_{\pi_{i}})\mbp(\hat{G}_{\pi_{i}})$
   \STATE $s_i = \min \big\{ 1, \exp(p_i - p_{i-1})\big\}$
   \STATE $z_i \sim \textrm{Bernoulli}(s_i)$
   \IF{$z_i = 0$}
   \STATE $\pi_{i} = \pi_{i-1}$ and $\hat{G}_{\pi_i}$ = $\hat{G}_{\pi_{i-1}}$ (chain does not move)
   \ENDIF
   \IF{$T$ is divisible by $\ceil{\frac{1}{\kappa}}$}
   \STATE Store $\hat{G}_{\pi_i}$
   \ENDIF
   \ENDFOR   
\end{algorithmic}
\label{algo}
\end{algorithm}

\section{Experiments} \label{sec:experiments}

In this section we empirically compare minimal I-MAP MCMC to order and partition MCMC. We~chose~\mbox{partition} MCMC since it does not have the bias of order MCMC and empirically has faster mixing than structure MCMC \cite{partition_mcmc}. We use the max-min-hill-climbing  (MMHC) algorithm \cite{mmhc} in conjunction with the nonparametric DAG bootstrap approach \cite{bootstrap_dag} as an additional baseline for comparison. For each dataset, we ran the Markov chains for $10^5$ iterations, including a burn-in of $2 \times 10^4$ iterations, 
and thinned the remaining iterations by a factor of 100. Seeded runs correspond to starting the Markov chain at the permutation/DAG obtained using MMHC. We also considered "cheat" runs that start at the true permutation with the intuition that we expect high scores on the true generating model. In terms of software, we used the code provided by \citet{partition_mcmc} to run partition and order MCMC. We used the method and software of \citet{count_lin_ext} for counting linear extensions for bias correction, and we implemented minimal I-MAP MCMC using the R-package \texttt{bnlearn}. 

\subsection{Prior and Likelihood} \label{sec:prior_lik}

As in many applications, a prior that induces sparsity in the underlying structure is desirable for interpretability and computation. Further, note that the true DAG $G^{*}$ is equal to the sparsest minimal I-MAP $G_{\pi}^*$ over all permutations $S_p$ based on CI relations $\mathcal{O}^{*}$  \cite{verma_pearl, greedy_sp, sp_algo}; thus, on minimal I-MAP space, a sparsity prior is natural. To achieve this end, we choose a prior of the form
$$
	\mbp(G) = \mbp^{*}(G) \exp \big(-\gamma \Vert G \Vert \big),
$$
where $\mbp^{*}(G)$ can include any structural information known about the DAG. Except where explicitly mentioned in what follows, we use this prior with $\mbp^{*}(G)$ uniform over DAGs. We note that, unlike our method or partition MCMC, order MCMC uses a uniform prior over permutations; the induced prior over DAGs as a result of such a prior is $\mbp_{\mathsf{order}}(G) = |\# \mathsf{linext}(G)|\mbp(G)$, where $|\# \mathsf{linext}(G)|$ denotes the number of linear extensions of $G$ \cite{ellis_wong}. Finally, each method assumes the data follow a multivariate Gaussian distribution with a Wishart prior on the network parameters. This assumption allows computation of $\mathbb{P}(D | G)$ via the \textit{BGe} score \cite{bge_orig, bge_score}.

\subsection{Mixing and Convergence}
We consider three different datasets. The first two were obtained by simulating data from a network consisting of $p=30$ nodes with $n=100$ and $n=1000$ observations respectively. The data were generated according to a linear structural equation model with additive Gaussian noise, where the edge weights on the underlying DAG $G^*$ were sampled uniformly from $[-1, -.25] \cup [.25, 1]$ as in \cite{greedy_sp}. The third dataset is from the \textit{Dream4} in-silico network challenge \cite{dream4_data} on gene regulation. In particular, we examine the \textit{multifactorial} dataset consisting of ten nodes and ten observations. 
\begin{figure}
\begin{centering}
\includegraphics[width=\linewidth]{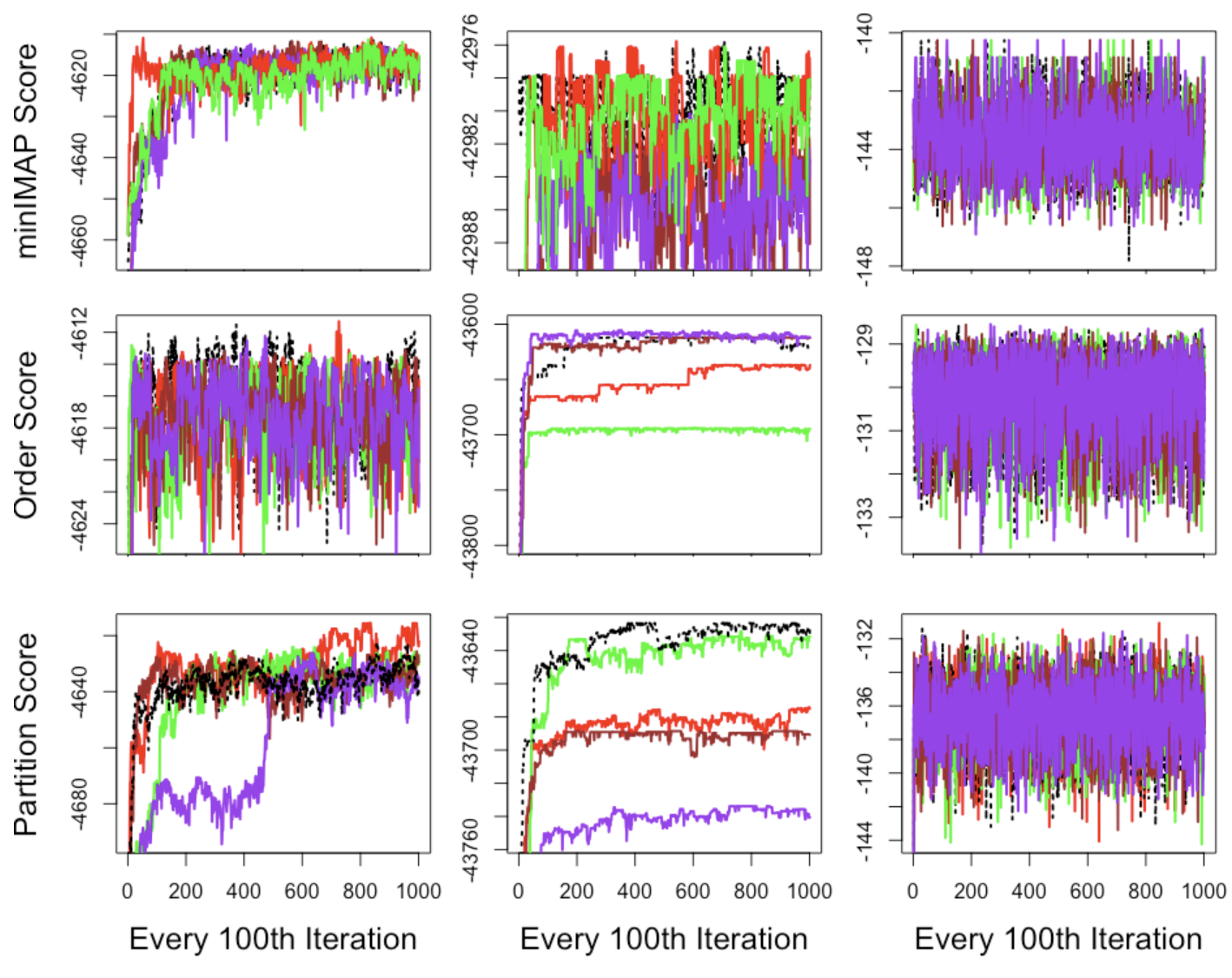}\par
\end{centering}
\vspace{-.1in}
\caption{From left to right, the columns represent the $n=100$, $n=1000$, and Dream4 datasets, respectively. From top to bottom, the rows correspond to minimal I-MAP (minIMAP), order, and partition MCMC. The black dotted line corresponds to runs seeded with the true permutation. The purple and brown lines correspond to runs seeded with a random permutation and the red and green curves represent runs seeded with MMHC.}
\label{fig:mcmc_score_runs}
\vspace{-.1in}
\end{figure}

In Figure \ref{fig:mcmc_score_runs} we analyze the mixing performance of the different methods. The convergence of different runs to the same score neighborhood can be taken as an indication of adequate mixing. Figure \ref{fig:mcmc_score_runs} suggests that for the $n=100$ and the Dream4 dataset all three methods have mixed well while for the dataset with $n=1000$ samples there is evidence of poor mixing in all methods since the posterior landscape is more peaky due to increased sample size. However, the score plots are markedly worse for order and partition MCMC. 

\begin{figure}
\begin{centering}
\includegraphics[width=\linewidth]{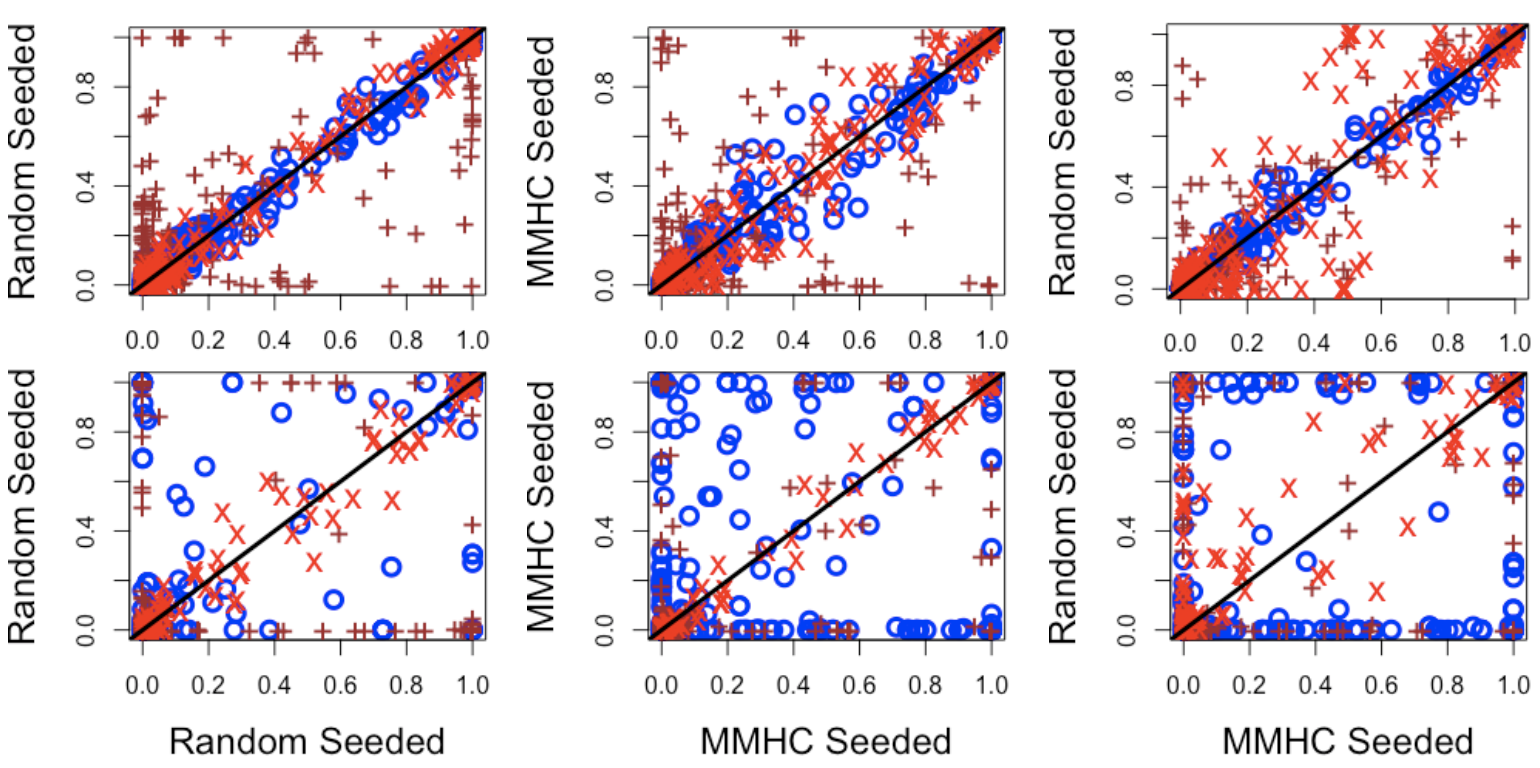}\par
\end{centering}
\vspace{-.1in}
\caption{The top row and bottom row correspond to the $n=100$ and $n=1000$ datasets respectively. Each point represents the posterior probability of a directed feature estimated by different seeded runs of MCMC. We consider all possible combinations of random and MMHC seeded runs for completeness. Red (x), blue (o), and brown (+) correspond to minIMAP, order, and partition MCMC respectively.}
\label{fig:seed_correlation}
\vspace{-.1in}
\end{figure}

Since the end goal is often to obtain robust estimates of particular feature probabilities, in Figure \ref{fig:seed_correlation} we analyze the correlation between the approximated posterior probabilities of directed features with respect to different seeded runs. Figure \ref{fig:seed_correlation} shows that the correlation is higher across all models for the dataset with $n=100$ samples, which is to be expected, since the chains seem to have mixed given the analysis in the score plots in Figure~\ref{fig:mcmc_score_runs}. Conversely, for the $n=1000$ dataset, Figure~\ref{fig:seed_correlation} shows that order and partition MCMC yield vastly different posterior probabilities across different runs, while minimal I-MAP MCMC maintains high correlation, thus suggesting again superior mixing.  

\subsection{ROC Performance} \label{sec:roc_perform}

As described in Section \ref{prelims}, the bias of order MCMC can be removed by dividing the functional of interest $f(G_t)$ by the number of linear extensions of $G_t$, where $G_t$ is a DAG sampled during the Monte-Carlo Step. We denote this by \emph{full bias correction} (\emph{FBC}). Although this leads to an unbiased estimator for order MCMC, there is a bias-variance trade-off. If a sampled DAG has few linear extensions, this DAG will be given more weight in the Monte Carlo step, thereby increasing the variance. Therefore, we also consider a \emph{partial bias correction} (\emph{PBC}), where the weights are truncated and re-normalized to belong to the 25th and 75th quartile of the inverse linear extension counts of the sampled DAGs. Finally, we denote \emph{no bias correction} by \emph{NBC}. 

In Table \ref{AUROC_vals} we report the area under the ROC curves (AUROC) for detecting directed and undirected features for the different methods. For order MCMC, we see a marginal performance boost after bias correction on the simulated datasets, but worse performance on Dream4. For the $n=100$ and $n=1000$ datasets, the Bayesian models perform better than the MMHC bootstrap. While Table \ref{AUROC_vals} shows that MMHC achieves the highest AUROC performance on the Dream4 dataset, the corresponding ROC plot provided in Figure \ref{fig:roc_plots} in Appendix \ref{a:roc_plots} shows that minimal I-MAP MCMC and order MCMC compare favorably to MMHC when the true negative rate (TNR), which equals one minus the false positive rate (FPR), is greater than 0.4. This range for the TNR is the relevant regime for biological applications, where it is often more important to control for Type I errors (i.e. incorrectly specifying causal relationships between nodes). 

The second column of Table \ref{AUROC_vals} for each dataset shows AUROC performance on the compelled edges and Figure \ref{fig:roc_plots} in Appendix \ref{a:roc_plots} contains the corresponding ROC plots. Recovering compelled edges is important because these are the only causal effects that are identifiable from observational data alone \cite{pearl_causality}.
Table \ref{AUROC_vals} shows that minimal I-MAP MCMC achieves the best performance in terms of recovering compelled edges on the $n=1000$ dataset and is marginally better than the other methods on the $n=100$ dataset.  
\begin{table}[!t]
\caption{AUROC results by dataset and method. NBC, PBC, and FBC stand for no, partial, and full bias correction. The columns represent AUROC values for undirected and compelled features respectively.}
\vspace{-.1in}
\vskip 0.15in
\begin{center}
\begin{small}
\begin{sc}
\begin{tabular}{lcccr}
\toprule
Method & n=100 & n=1000 & Dream4 \\
\midrule
minIMAP    &.946 \ \textbf{.695} & \textbf{1.00} \ \textbf{.958} & .574 \ .556 \\
Order-NBC    &\textbf{.957} \ .675 & .949 \ .395 &  $\mathbf{.599}$ \ .600     \\
Order-PBC     &.956 \ .677& .949 \ .393 & .579 \ .444  \\
Order-FBC       &.952 \ \textbf{.695} &.950 \ .395 & .563 \ .489  \\
Partition      & .857 \ .660 &.890 \ .674 & .497 \ \textbf{.733}       \\
MMHC-Boot    & .842 \ .693 & .892 \  .668& .552 \ .533 \\
\bottomrule
\end{tabular}
\end{sc}
\end{small}
\end{center}
\vskip -0.1in
\label{AUROC_vals}
\vspace{-.1in}
\end{table}
\subsection{Time and Memory Complexity}
\begin{figure}
\begin{centering}
\includegraphics[width=.78\linewidth]{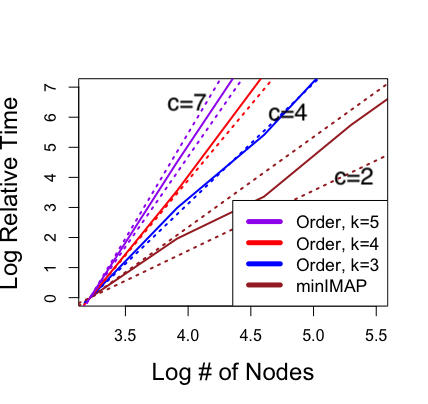}\par
\end{centering}
\vspace{-.1in}
\caption{Average iteration times for different sized networks. The times are relative to the average iteration time for $p=25$ nodes; $c$ denotes the slope of the dotted lines and estimates the computational complexity $\mathcal{O}(p^c)$.}
\label{speed_complexity}
\vspace{-.1in}
\end{figure}

Since partition MCMC has a similar time and memory complexity as order MCMC, we focus on comparing minimal I-MAP MCMC to order MCMC in these regards. Recall that $p$ denotes the number of nodes and $k$ denotes the maximum indegree of the underlying DAG $G^*$. To control for different implementations, we computed the average iteration times relative to the average iteration time for $p=25$ nodes. The average iteration times do not include the time it takes to cache all the scores in order MCMC and the time it takes to construct $\hat{G}_{\pi_0}$ for initiating the Markov chain. Figure \ref{speed_complexity} shows that order MCMC scales similarly to its predicted theoretical complexity of $\mathcal{O}(p^{k+1})$. For minimal I-MAP MCMC, we provided a bound of $\mathcal{O}(p^{4})$ in Proposition~\ref{prop:time_memory_ergodic}. Figure \ref{speed_complexity} suggests that the complexity scales by a factor of $p$ better than the bound we obtained, namely $\mathcal{O}(p^{3})$. Finally, we note that order MCMC runs out of memory quickly when either $k$ or $p$ grows. As a specific example, for only $p=80$ nodes and $k=5$, order MCMC takes over 40 GB of space while minimal I-MAP MCMC takes around 1 MB.      
\subsection{Incorporating Priors} \label{path_order_priors}
Unlike minimal I-MAP MCMC, both partition and order MCMC require that the prior over DAGs  factorizes as $\mathbb{P}(G) = \prod_{i=1}^p \rho(X_i, \mathsf{Pa}_{G}(X_i))$ which is defined as \textit{structure modularity} by \citet{Friedman2003}. $\mathbb{P}(G)$ is used in order MCMC to specify the conditional distribution $\mbp(G \mid \pi) = I(G \precsim \pi)\mathbb{P}(G)$, which is needed to calculate the likelihood $\mbp(D \mid \pi)$ in order MCMC; see also Appendix \ref{dis:priors_topo_diff}. 
The assumption of structure modularity is a practical limitation. In biological applications, for example, prior information often comes in the form of path information between classes of vertices, which is not structure modular in general. In the following, we illustrate this point using the biological network studied by \citet{speed_priors} (reproduced in Figure \ref{speed_net_roc} in Appendix \ref{a:roc_plots}). In this application, we have  prior knowledge on both orders and paths. In particular, we expect ligands to come before receptors, and receptors before cytosolic proteins. In addition, we expect to see paths from ligands to receptors and paths from receptors to cytosolic proteins \cite{speed_priors}. Such path information cannot be used by order and partition MCMC since this information is not structure modular. To test if path knowledge leads to better inference, we compared the ROC plots (Figure~\ref{speed_net_roc}, Appendix \ref{a:roc_plots}) and AUROC (Table \ref{prior_auc}) for directed edge recovery for the different methods. Table~\ref{prior_auc} shows that the path prior leads to a boost in AUROC performance of minimal I-MAP MCMC by $1-2\%$ percent,  thereby suggesting that structure modularity can be limiting for certain applications. The specific form of the path and order prior are provided in Appendix~\ref{dis:path_order_prior}. 

\begin{table}[t] 
\caption{AUROC results for directed edge recovery in the protein network in Figure \ref{speed_net_roc}.}
\vskip 0.15in
\begin{center}
\begin{small}
\begin{sc}
\begin{tabular}{lcccr}
\toprule
Method & AUROC \\
\midrule
minIMAP w/ path and order prior   &\textbf{.929} \\
minIMAP w/ order prior   & .917 \\
Order w/ order prior & .874 \\
Partition w/ order prior   & .912 \\
MMHC-Boot & .909 \\
\bottomrule
\end{tabular}
\end{sc}
\end{small}
\end{center}
\vskip -0.1in
\label{prior_auc}
\end{table}

\section{Concluding Remarks}
In this paper, we introduced minimal I-MAP MCMC, a new Bayesian approach to structure recovery in causal DAG models. Our algorithm works on the data-driven space of minimal I-MAPs with theoretical guarantees on posterior approximation quality. We showed that unlike order or partition MCMC the complexity of an iteration in minimal I-MAP MCMC does not depend on the maximum indegree of the true underlying DAG. This theoretical result was confirmed in our empirical study. In addition, our empirical study showed that minimal I-MAP MCMC achieves similar or faster mixing than other state-of-the-art methods for Bayesian structure recovery. 

While we have focused on the Gaussian setting, it would be interesting in future work to extend the theoretical analysis to other distributions, in particular the discrete setting. Finally, it would be interesting to explore the performance of minimal I-MAP MCMC for obtaining MAP estimates or as a new DAG scoring criterion. In particular, the scoring criterion of the \textit{greedy SP (GSP)} algorithm \cite{greedy_sp} is equivalent to our DAG score (i.e., unnormalized posterior probability) when $\gamma \rightarrow \infty$ in the prior in Section \ref{sec:prior_lik} and the search space is restricted to $\hat{\mathcal{G}}$. In this case, the likelihood term has no influence in picking the minimal I-MAP from $\hat{\mathcal{G}}$. We might therefore find improved performance in terms of structure recovery over the GSP algorithm by incorporating the likelihood term by setting $\gamma < \infty$.

\section*{Acknowledgements}  Raj Agrawal was supported by an MIT Aziz Ashar Presidential Fellowship. Tamara Broderick was supported in part by an ARO Young Investigator Program Award, ONR grant N00014-17-1-2072, and a Google Faculty Research Award. Caroline Uhler was supported in part by NSF (DMS-1651995), ONR (N00014-17-1-2147), and a Sloan Fellowship.


\bibliography{example_paper}
\bibliographystyle{icml2018}

\clearpage

\appendix

\section{CI Testing for Gaussian Data} \label{A:ci_tests}

In the case of multivariate Gaussian data, one may use the Fisher z-transform \citep{fisherz} to perform CI testing. The Fisher z-transform is given by  
$$Z(i, j \mid S) \defeq \frac{1}{2}\frac{\log(1 + \hat{\rho}_{i,j\mid S})}{\log(1 - \hat{\rho}_{i,j\mid S})}, $$
where $\hat{\rho}_{i,j\mid S}$ is the empirical partial correlation between $X_i$ and $X_j$ given $X_S$. To conduct a two-sided hypothesis test at significance level $\alpha$, one may test if 
$$\sqrt{n - \abs{S} - 3}|Z(i, j \mid S)| \leq \Phi^{-1}(1 - \alpha / 2),$$
where $\Phi^{-1}$ is the inverse CDF of $N(0, 1)$.

\section{Update Algorithm} \label{A:update_algo}
Algorithm \ref{update_min_imap} specifies the update procedure used in Algorithm \ref{algo} to reduce the number of CI tests needed.
\begin{algorithm}[tb] 
   \caption{Update Minimal I-MAP (\emph{UMI})}
\begin{algorithmic}
   \STATE {\bfseries Input:} Current permutation $\pi_i$, previous permutation $\pi_{i-1}$, previous minimal I-MAP $G_{\pi_{i-1}}$, significance level $\alpha$, data $D$
   \STATE {\bfseries Output:} $G_{\pi_{i}}$
   \STATE $k$ = min index of adjacent transposition
   \IF{$k = 1$ (first and last element swapped)}
   \STATE Compute $\hat{G}_{\pi_i}$ from $\hat{\mathcal{O}}_n(D,\alpha)$
   \ELSE
   \STATE $G_{\pi_{i}}$ = $G_{\pi_{i-1}}$
   \STATE Reverse edge from $X_{\pi_{i}(k+1)}$ to $X_{\pi_{i}(k)}$ in $G_{\pi_{i}}$ if such an edge exists
   \FOR{$s=1$ {\bfseries to} $k-1$} 
   \FOR{$j=k$ {\bfseries to} $k+1$}
   \STATE $S = \{\pi(1), \cdots, \pi(j - 1) \} \setminus \{\pi(s) \}$
   \STATE Let $z = \hat{\mathcal{O}}^{(n)}_{i,j|S}(D,\alpha)$ 
   \STATE Update edge from $X_{\pi(s)}$ to $X_{\pi(j)}$ to $z$ in $G_{\pi_{i}}$
   \ENDFOR
   \ENDFOR  
   \ENDIF
\end{algorithmic}
\label{update_min_imap}
\end{algorithm}

\section{Discussion of the Assumptions} \label{dis_of_assump}


Based on the discussion of \citet{high_dim_PC}, Assumption \ref{assumptions}(b) is not such a strong assumption and seems more of a regularity condition needed to prove the bounds. Assumption \ref{assumptions}(d) has an intuitive interpretation; it says that the best prediction of $G$ based on the data and order is captured by the constructed network. Conditioned on the order, the inference problem is not hard; i.e., we just need to recover the skeleton. Since we can recover the skeleton via the empirical CI relations, $\hat{G}_{\pi}$ is indeed the best prediction of the network given the data and order in many cases, which would imply that $\hat{G}_{\pi}$ can reasonably be assumed to be a sufficient statistic. Assumption \ref{assumptions}(e) is a quite weak assumption; it says that the information of $\hat{G}_{\pi}$ does not help in predicting the probability of a CI error. This makes sense because we want to know if $\hat{G}_{\pi}$ does not equal $G_{\pi}^*$. But, without observing $G_{\pi}^*$, or conditioning on some property of $G_{\pi}^*$ in addition to $\hat{G}_{\pi}$, it seems reasonable to assume that our prediction is left unchanged when knowing $\hat{G}_{\pi}$.

\section{Proofs}
\subsection{Proof of Lemma \ref{thm:exp_concentrationv2}} \label{pf:exp_concentrationv2}
The proof relies heavily on the concentration bounds used to prove the high-dimensional consistency of the PC algorithm \cite{high_dim_PC}. To start, notice that 
\begin{align}
	\nonumber
\mbp(G_{\pi} \neq \hat{G}_{\pi} \mid G, \theta)
                 & = \mbp(\textrm{CI error(s) constructing $\hat{G}_{\pi}$})   \\  
                 \label{union_bound}
                 & \leq \sum_{i=1}^{p-1} \sum_{j=i+1}^{p} \mbp(E_{i,j}(G^{*}, \theta^{*})), 
\end{align}
where $E_{i, j}(G^{*}, \theta^{*})$ is the event that a CI error is made when testing $X_{\pi(i)} \nindep X_{\pi(j)} \ | \ X_S$, for $S = \{\pi(1), \cdots, \pi(j-1) \} \setminus \{\pi(i) \}$, conditioned on the Bayesian network $(G^{*}, \theta^{*})$ generating the observed data. Note that these tests are performed at the significance level provided in the statement of the lemma. 

By assumption, $Q^{*}_{\theta^{*}, G^{*}} \leq q^{*} < 1$ and $0 < r^{*} \leq R^{*}_{\theta^{*}, G^{*}}$ (without loss of generality, ignore measure zero sets). Picking such $q^{*}$ and $r^{*}$ then satisfy the assumptions required in Lemma 4 of \citet{high_dim_PC}. Equations (16) and (17) from \citet{high_dim_PC} imply that there exist constants $C_1$, $C_2$ that depend \textit{only} on $q^*$ such that  
$$\mbp({E_{i, j}}(G^{*}, \theta^{*})) \leq C_1(n - p)\exp \{{-C_2(r^{*})^2(n - p)} \}$$
for any $i,j$. Hence, 
\begin{equation} \label{concen_cond_on_dag}
\mbp(G_{\pi}^{*} \neq \hat{G}_{\pi} \mid G, \theta) \leq f(n,p),
\end{equation}
where $f(n, p) = p^2 C_1(n - p)\exp \{-C_2(r^{*})^2(n - p)\}$.

Now, 
\begin{align*}
	\lefteqn{\mbp(G_{\pi}^{*} \neq \hat{G}_{\pi})} \\
    	&= \sum_{G \in \mathcal{G}} \int_{\theta} \mbp(G_{\pi}^{*} \neq \hat{G}_{\pi} \mid G, \theta)\mbp(\theta \mid G) \mbp(G)d\theta \\
        &\leq \sum_{G \in \mathcal{G}} \int_{\theta} f(n,p)\mbp(\theta \mid G) \mbp(G) d\theta \\
        &= f(n,p),
\end{align*}
as desired. 

\subsection{Proof of Theorem \ref{thm:good_approx_feats}} \label{pf:good_approx_feats}
By the tower property, 
$$\mbe_{\mbp(G \mid D)} f(G) = \mbe_{\mbp(\pi \mid D)} \mbe_{\mbp(G \mid D, \pi)} f(G).$$

As before, define $A_{\pi}$ as the event that $\hat{G}_{\pi} = G_{\pi}^{*}$. 
We may expand $\mbe_{\mbp(G \mid D, \pi)} f(G)$ as  
\begin{align*}
	\lefteqn{\mbe_{\mbp(G \mid D, \pi)} f(G)} \\
    	&= \sum_{G \in \mathcal{G}} f(G) \mbp(G \mid D, \pi) \\
        &= \sum_{G \in \mathcal{G}} f(G) \mbp(G, A_{\pi} \mid \hat{G}_{\pi}) +  \sum_{G \in \mathcal{G}} f(G) \mbp(G, A_{\pi}^C \mid \hat{G}_{\pi}) \\
        &\quad \textrm{by Assumption \ref{assumptions}(d) and the law of total probability} \\
        &= f(\hat{G}_{\pi}) + \mbp(A_{\pi}^C \mid \hat{G}_{\pi}) \\
        & \quad \cdot \left[\sum_{G \in \mathcal{G}} f(G) \mbp(G \mid \hat{G}_{\pi}, A_{\pi}^C) - f(\hat{G}_{\pi}) \right] \\
        &\quad  \textrm{by the fact that $\mbp(G \mid \hat{G}_{\pi}, A_{\pi}) = I(G = \hat{G}_\pi)$} \\
        &\quad \textrm{according to the exact reasoning used in Section \ref{gen_model}} \\
        &= f(\hat{G}_{\pi}) + \mbp(A_{\pi}^C) \\
        & \quad \cdot \left[\sum_{G \in \mathcal{G}} f(G) \mbp(G \mid \hat{G}_{\pi}, A_{\pi}^C) - f(\hat{G}_{\pi}) \right],
\end{align*}
where the final equality uses Assumption \ref{assumptions}(e).

We claim that
\begin{equation}
\mbe_{\mbp(\pi \mid D)} f(\hat{G}_{\pi}) = \mbe_{\hat{\mbp}(G \mid D)} f(G).
\label{eqn:expect_under_phat}
\end{equation}
To prove Equation~(\ref{eqn:expect_under_phat}), notice that
\begin{align*}
	\lefteqn{\mbe_{\mbp(\pi \mid D)} f(\hat{G}_{\pi})} \\
        &= \sum_{\pi \in S_{p}} f(\hat{G}_{\pi}) \mbp(\pi \mid D) \\
        &= \sum_{G \in \mathcal{G}} f(G) \sum_{\pi \in S_{p}} \mathbbm{1}\{G \in \hat{\mathcal{G}}\} \mathbbm{1}\{ G = \hat{G}_{\pi}\} \mbp(\pi \mid D) \\
        &= \sum_{G \in \mathcal{G}} f(G) \hat{\mbp}(G \mid D) \\
        &= \mbe_{\hat{\mbp}(G \mid D)} [f(G)].
\end{align*}
Finally,  
$$\left|\sum_{G \in \mathcal{G}} f(G) \mbp(G \mid \hat{G}_{\pi}, A_{\pi}^C) - f(\hat{G}_{\pi}) \right| \leq 2M$$
and 
$$\mbp(A_{\pi}^c) \leq C_1(n - p)\exp \{{-C_2(r^{*})^2(n - p)} \}$$
by Lemma \ref{thm:exp_concentrationv2}. The result now follows by taking expectations and using the above bounds.

\subsection{Proof of Proposition \ref{prop:stationary_distr}} \label{pf:stat_dist}

Ergodicity follows from the fact that any permutation can be reached from adjacent transpositions, and aperiodicity follows from our constraint that $s \in (0, 1)$. Since adjacent transpositions trivially satisfy the detailed balance equations, the Markov chain has stationary distribution $\mbp(\hat{G}_{\pi} \mid D)$.

\subsection{Proof of Proposition \ref{prop:cache_update}} \label{pf:cache_update}
\begin{prop} \label{prop:cache_update}
If $\pi_t$ and $\pi_{t+1}$ differ by an adjacent transposition, Algorithm \ref{update_min_imap} correctly calculates $\hat{G}_{\pi_{t+1}}$ from $\hat{G}_{\pi_{t}}$.  
\end{prop}

This update rule was also used by \citet{greedy_sp}. We here provide the proof for completeness. The result trivially follows if $\pi_{t+1}$ is obtained by swapping the first and last element of $\pi_t$ since all CI tests are recomputed in this case. Hence, we may assume $\pi_t$ and $\pi_{t+1}$ differ by an adjacent transposition not at the border.  Suppose $\pi_t = (n_1 \cdots n_i n_{i+1} \cdots n_p)$ and $\pi_{t+1} = (n_1 \cdots n_{i+1} n_i \cdots n_p)$, where the permutations differ at an adjacent permutation at position $i$. Then, the only edges that can be different in $\hat{G}_{\pi_{t}}$ and $\hat{G}_{\pi_{t+1}}$ are those edges connected nodes $n_i$ / $n_{i+1}$ with nodes $n_k, 1\leq k < i$. 
Correcting the edges $(n_{i}, n_k)$ and $(n_{i+1}, n_k)$ corresponds to recomputing the conditional independence statements $X_{n_{i}} \nindep X_{n_k} \mid X_{S_{i}}$ and $X_{n_{i+1}} \nindep X_{n_k} \mid X_{S_{i+1}}$, for $X_{S_{i}} = \{n_1, \cdots, n_{i+1}\} \setminus \{n_k\}$ and $X_{S_{i+1}} = \{n_1, \cdots, n_{i-1}\} \setminus \{n_k\}$ and updating the corresponding edges. The for loop in Algorithm \ref{update_min_imap} carries out the CI tests specified in the previous sentence. Finally, we need to reverse the edge between nodes $X_{n_{i}}$ and $X_{n_{i+1}}$ if there was an edge between them in the old DAG $\hat{G}_{\pi_{t}}$; this reversal is accomplished at the very start of Algorithm \ref{update_min_imap}.

\subsection{Proof of Proposition \ref{prop:time_memory_ergodic}} \label{pf:time_mem_scale}
The memory complexity follows trivially from the fact that it takes $\mathcal{O}(p^2)$ memory to store $\hat{G}_{\pi}$ in an adjacency matrix. Computing partial correlations takes at most $\mathcal{O}(p^3)$ time using the well-known partial correlation recursive formula \cite{part_cor_form}. Instantiating $\hat{G}_{\pi_0}$ requires $\mathcal{O}(p^2)$ CI tests and hence takes at most $\mathcal{O}(p^5)$ time to compute. The subsequent $\hat{G}_{\pi_i}$ are computed using Algorithm \ref{update_min_imap}. The correctness of Algorithm \ref{update_min_imap} was shown in Appendix \ref{pf:cache_update}. We claim Algorithm \ref{update_min_imap} takes average case $\mathcal{O}(p^4)$ time. 

First, we show that the first and last elements of $\pi_i$ are swapped with probability less than $\frac{1}{p}$ when moving from $\pi_i$ to $\pi_{i+1}$. Notice from our definition of the adjacent transposition distribution $q$ that the probability of either the first or last element undergoing an adjacent transposition is $\frac{2(1-s)}{p}$. Conditioned on either the first or last element being chosen to be swapped, there is probability $\frac{1}{2}$ that the first (last) element will be swapped with the last (first) element. Hence, the probability of the first and last element being swapped equals $\frac{(1-s)}{p}$ which is less than $\frac{1}{p}$. When the first and last element are swapped, all $p^2$ CI tests need to be recomputed. All the remaining adjacent transpositions require at most $2p$ additional CI tests to be performed in the for loop of Algorithm \ref{update_min_imap}. Hence, on average, the number of additional CI tests is $\mathcal{O}(p)$ which implies the average running time of Algorithm \ref{update_min_imap} is $\mathcal{O}(p^4)$.

\section{Justification for restricting the prior space} \label{dis:prior_red}
Following nearly the same reasoning used to motivate our likelihood approximation in Section \ref{gen_model}, here we justify 
$$
\mbp( \pi \mid \hat{\mathcal{O}}_{n})
	\approx \mbp(G = \hat{G}_{\pi}).
$$
Notice that
\begin{align*}
	\lefteqn{ \mbp( \pi \mid \hat{\mathcal{O}}_{n}) } \\
     &= \mbp( \pi \mid \hat{\mathcal{O}}_{n}, A_{\pi}) \mbp(A_{\pi} | \hat{\mathcal{O}}_{n}) + \mbp( \pi \mid \hat{\mathcal{O}}_{n}, A^{C}_{\pi}) \mbp(A^{C}_{\pi} | \hat{\mathcal{O}}_{n}) \\
     &= \mbp( \pi \mid \hat{\mathcal{O}}_{n}, A_{\pi}) \mbp(A_{\pi}) + \mbp( \pi \mid \hat{\mathcal{O}}_{n}, A^{C}_{\pi}) \mbp(A^{C}_{\pi}),
\end{align*}
where the final equality follows from Assumption \ref{assumptions}(e). We claim that
\begin{equation} \label{eq:prior_reduction}
\mbp( \pi \mid \hat{\mathcal{O}}_{n}, A_{\pi}) = \mbp(G = \hat{G}_{\pi}).
\end{equation}
Given $\mathcal{O}_{n}$, we can construct $\hat{G}_{\pi}$, and conditioned on $A_{\pi}$, $\hat{G}_{\pi} = G_{\pi}^*$. Each permutation $\pi$ may therefore be associated with its true corresponding DAG $G_{\pi}^*$ which equals $\hat{G}_{\pi}$. Hence, the conditional probability $\mbp(\pi | \hat{\mathcal{O}}_{n}, A_{\pi})$ equals the prior probability of $\hat{G}_{\pi}$, namely $\mbp(G = \hat{G}_{\pi})$. 

Finally, since $\mbp(A_{\pi})$ goes to zero exponentially fast by Lemma \ref{thm:exp_concentrationv2}, $\mbp( \pi | \hat{\mathcal{O}}_{n})$ is well approximated by $\mbp(G = \hat{G}_{\pi})$.


\section{Prior Specification on Topological Orderings} \label{dis:priors_topo_diff}

Here we illustrate the computational difficulty of specifying a posterior $\mbp(\pi | D)$ that agrees with our original prior $\mbp^*(G)$ and likelihood $\mbp(G | D)$ on the space of DAGs. Notice that
\begin{equation} \label{eq:order_lik}
\mbp(D \mid \pi) = \sum_{G} \mbp(D \mid G) \mbp(G \mid \pi).
\end{equation}
Equation~(\ref{eq:order_lik}) implies that we must specify a conditional distribution $\mbp(G | \pi)$ to calculate the likelihood term for $\mbp(\pi | D)$. To understand what this conditional distribution should be, notice that the induced prior over DAGs equals
\begin{equation} \label{induced_prior}
\mbp(G) = \sum_{\pi \in S_p} \mbp(G \mid \pi) \mbp(\pi).
\end{equation}
In order MCMC, the assumed prior $\mbp(\pi)$ is equal to $\frac{1}{p!}$ \cite{Friedman2003}. A natural distribution one may specify for $\mbp(G | \pi)$, and the one assumed in \cite{Friedman2003}, is
\begin{equation} \label{eq:bad_cond_dist}
\mbp(G \mid \pi) = I(G \precsim \pi) \mbp^*(G).
\end{equation}
However, it is trivial to check that Equation~(\ref{eq:bad_cond_dist}) implies Equation~(\ref{induced_prior}) equals $|\# \mathsf{linext}(G)|\mbp^*(G)$, where $|\# \mathsf{linext}(G)|$ denotes the number of linear extensions of $G$ \cite{ellis_wong}. Therefore, we instead need 
$$\mbp(G \mid \pi) = \frac{1}{|\# \mathsf{linext}(G)|}(G \precsim \pi) \mbp^*(G)$$
to construct a model that agrees with our desired prior $\mbp^*(G)$ on DAGs. The difficulty of defining a prior on $\mbp(\pi | \mathcal{O}_{n})$ is calculating $|\# \mathsf{linext}(G)|$, which is $\#P$ in general. We should note that we avoid these issues by instead defining a prior on $\mbp(\pi | \mathcal{O}_{n})$. $\mbp(\pi | \mathcal{O}_{n})$ allows us to define a distribution that \emph{approximately} induces the correct DAG prior; see the discussion in Section \ref{gen_model}.

\section{Path and Order Priors} \label{dis:path_order_prior}
Here we provide the specific form of the order and path priors used in the experiment in Section \ref{path_order_priors}. Let $L$, $R$, and $C$ denote the set of ligands, receptors, and cytosolic proteins, respectively, in the network in Figure \ref{speed_net_roc}. For the order prior, $\mbp(\pi)$, we set 
$$\mbp(\pi) \defeq \exp\bigg(\sum_{L}f_{L}(l) + \sum_{R}f_{R}(r)\bigg),$$
where $f_{L}(l)$ indicates if ligand node $l$ came before all nodes in $R \cup C$ and $f_{R}(r)$ indicates if receptor $r$ came before all nodes in $C$ and after $L$ in $\pi$. For our method, the order prior is incorporated into our prior on DAGs. Specifically, we replace the DAG prior of $\mbp(G) = \exp \big(-\gamma \Vert G \Vert \big)$ used in our other experiments with,
$$
\mbp(G) \defeq \exp \big(-\gamma \Vert G \Vert \big) \exp\bigg(\sum_{L}f_{L}(l) + \sum_{R}f_{R}(r)\bigg).
$$
%
We refer to the prior above as \textit{minIMAP w/ path prior} in Table \ref{table:auc_speed}. To incorporate path information, we take a prior of the form,  
$$\exp\bigg(\sum_{L}h_{L}(l) + \sum_{R}h_{R}(r)\bigg),$$
where $h_{L}(l)$ indicates if ligand node $l$ had a path to at least one node in $R$ and $h_{R}(r)$ indicates if receptor $r$ had a path to at least one node in $C$. Combined with the order prior, the prior \textit{minIMAP w/ path and order} in Table \ref{table:auc_speed} is given by,   

\begin{equation}
\begin{split}
\nonumber
\mbp(G) \defeq \exp \big(-\gamma \Vert G \Vert \big) \exp\bigg(\sum_{L}f_{L}(l) + \sum_{R}f_{R}(r)\bigg) & \\ \exp\bigg(\sum_{L}h_{L}(l) + \sum_{R}h_{R}(r)\bigg).
\end{split}
\end{equation}

\begin{figure*}
\begin{centering}
\includegraphics[width=\linewidth]{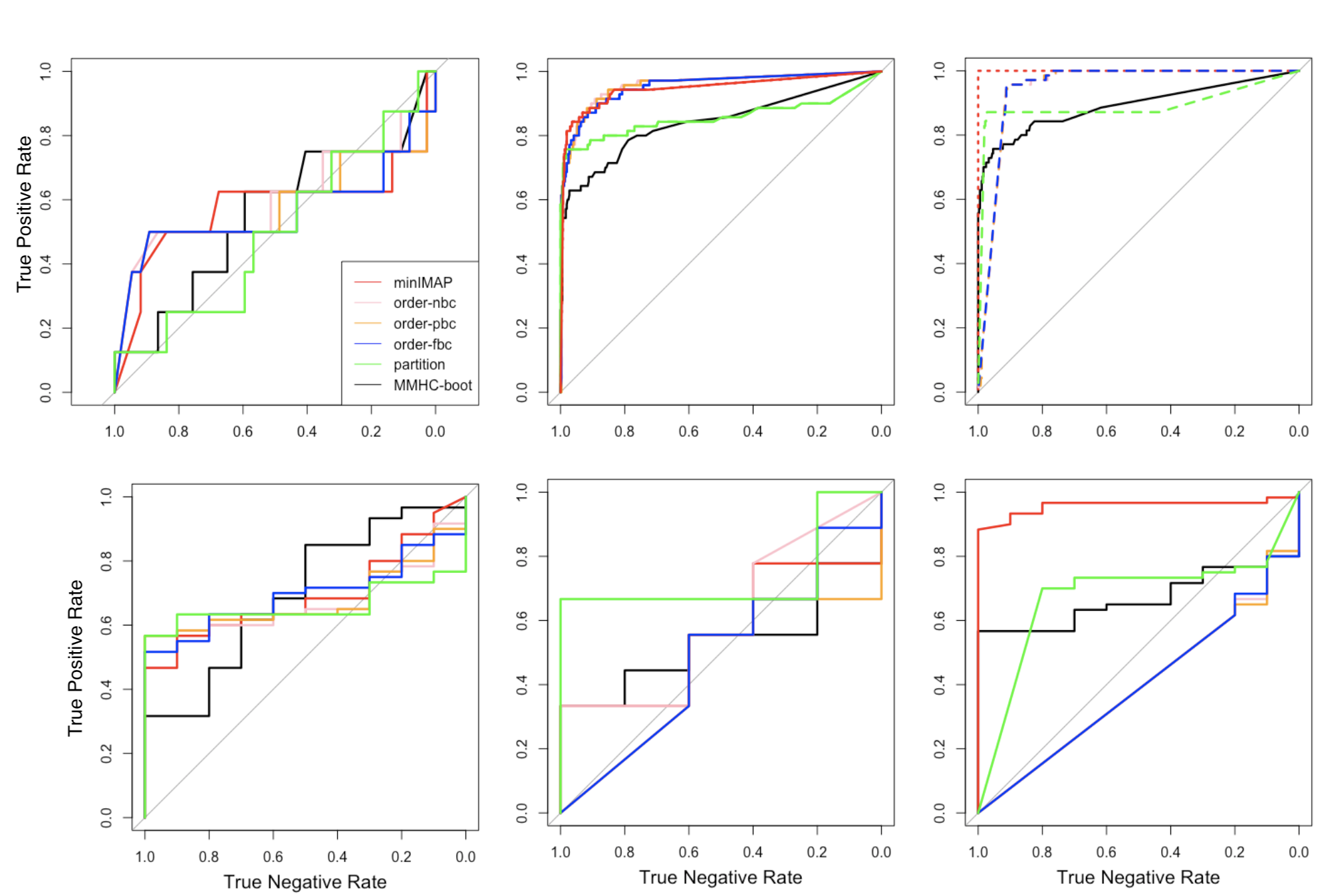}\par
\end{centering}
\vspace{-.1in}
\caption{The top ROC curves represent recovery of undirected features and the bottom for compelled features. From left to right, the plots correspond to the Dream4, n=100, and n=1000 datasets.}
\label{fig:roc_plots}
\vspace{-.1in}
\end{figure*}

\section{Additional Experiments and Plots} \label{a:roc_plots}

To further analyze the mixing behavior of the different methods, we compute the correlation between different seeded runs for estimating marginal directed edge probabilities. Table \ref{extra_sims} shows the average correlations and standard errors based on two hundred synthetic datasets with $n=1000$ observations and $p=30$ nodes. Note: Each method was run with $1 \times 10^5$ iterations and a burn-in of $2 \times 10^4$ iterations. 

\begin{figure*}[!t]
\begin{multicols}{2}
    \includegraphics[width=\linewidth]{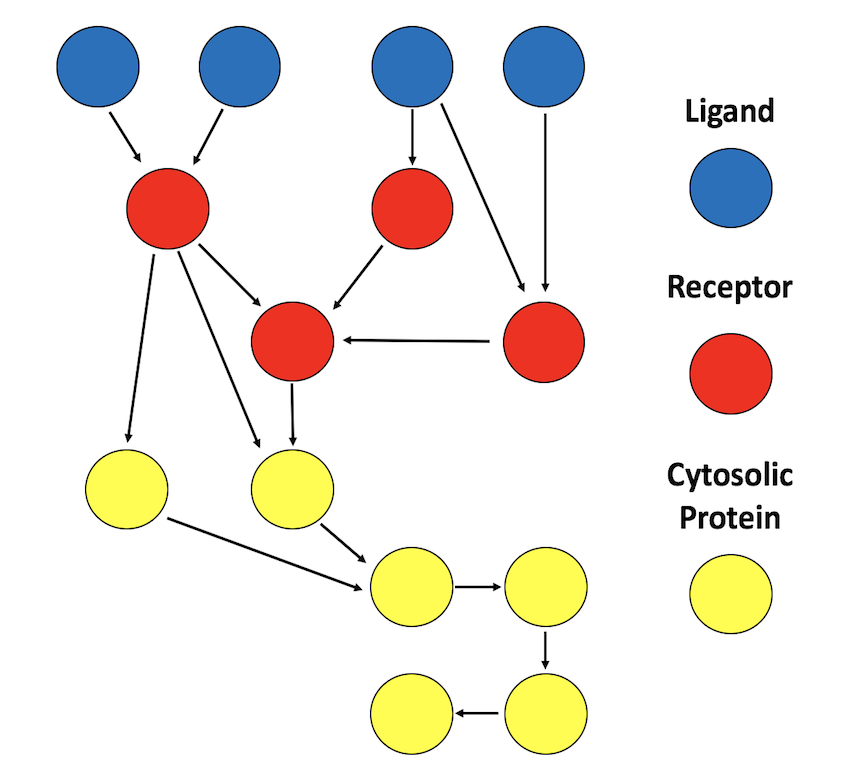}\par
    \includegraphics[width=1.1\linewidth]{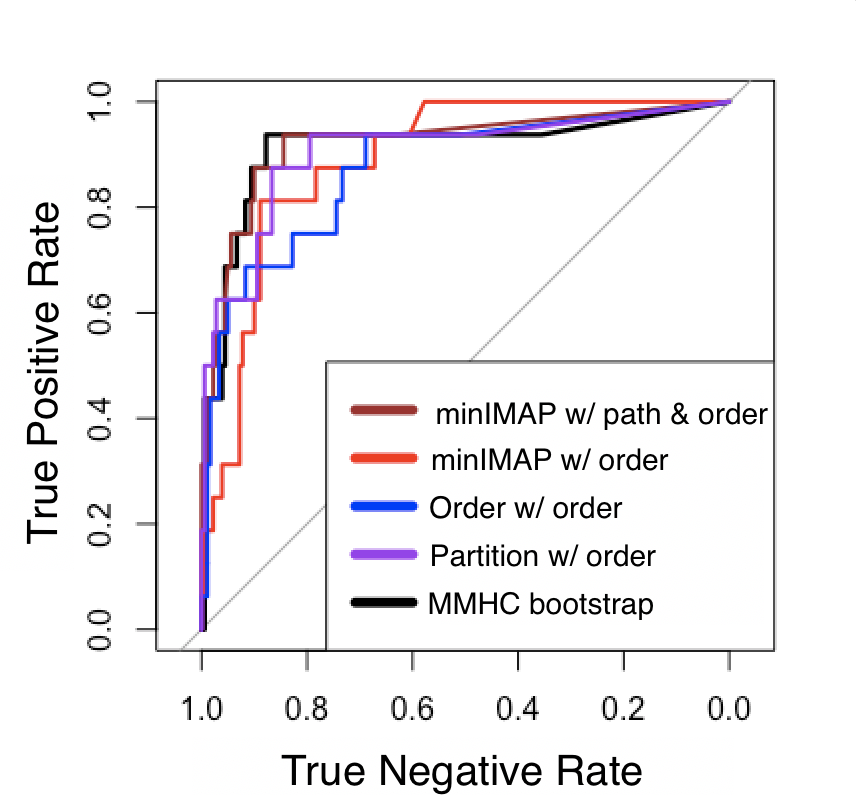}\par\
\end{multicols}
\vspace{-1cm}
\caption{The network on the left is taken from \cite{speed_priors}. The ROC plot on the right corresponds to the recovery of directed edges. Path and order refers to a prior that takes both path and order information into account as specified in Section \ref{path_order_priors}. For order and  partition MCMC, only order information can be used in the prior as discussed in Section \ref{path_order_priors}.}
\label{speed_net_roc}
\end{figure*}

The ROC plots for the $n=100$, $n=1000$, and Dream4 datasets are shown in Figure \ref{fig:roc_plots}; see Section \ref{sec:roc_perform} for a discussion of these plots. The network in \cite{speed_priors} used for the experiments in Section \ref{path_order_priors} is given in Figure \ref{speed_net_roc}. 

\begin{table}[t] \label{table:auc_speed}
\caption{Average correlation of directed features between runs seeded with the true network and runs seeded with MMHC from two hundred randomly generated DAGs with $p=30$ nodes. Higher is better.}
\vskip 0.15in
\begin{center}
\begin{small}
\begin{sc}
\begin{tabular}{lcccr}
\toprule
Method & Avg. Correlation & Std. Error \\
\midrule
minIMAP  &\textbf{.977} & .004 \\
Order & .928 & .007 \\
Partition & .784 & .006 \\
\bottomrule
\end{tabular}
\end{sc}
\end{small}
\end{center}
\vskip -0.1in
\label{extra_sims}
\end{table}


\end{document}